\documentclass[fleqn,10pt]{wlscirep}
\usepackage[utf8]{inputenc}
\usepackage[T1]{fontenc}
\title{A simple model for pink noise from amplitude modulations}

\author[1,*]{Masahiro Morikawa}
\author[2]{Akika Nakamichi}
\affil[1]{Department of Physics, Ochanomizu University ~\\
2-1-1 Otsuka, Bunkyo, Tokyo 112-8610, Japan}
\affil[2]{General Education, Kyoto-Sangyo University ~\\
Motoyama Kamigamo Kita-ku, Kyoto 603-8555 Japan}

\affil[*]{hiro@phys.ocha.ac.jp}


\begin{abstract}
We propose a simple model for the origin of pink noise (or 1/f fluctuation)
based on the beat of cooperative waves. These cooperative waves arise
spontaneously in a system with synchronization, resonance, and infrared
divergence. Many cooperative waves with close frequencies can produce
signals of arbitrary small frequencies from a system of small size.
This beat mechanism can be understood as amplitude modulation. The
pink noise can appear after the demodulation process, which produces
a variety of pink noise in many fields. The pink noise thus formed
from the beat has nothing to do with dissipation or long-time memory.
We also suggest new ways of looking at pink noise in shallow earthquakes,
solar flares, and stellar activities. 
\end{abstract}
\begin{document}

\flushbottom
\maketitle
%
%
\thispagestyle{empty}

\section{\label{sec:1}Introduction}

Pink noise is ubiquitous. This noise is characterized by the power-law
behavior in the very low-frequency region of the power spectrum density
(PSD) with power $-\alpha$, ($0.5 < \alpha <
1.5$). This noise is also known as 1/f fluctuation or flicker noise.

Since the first discovery of pink noise in a vacuum tube current \cite{Johnson1925},
the same noise has been observed in many systems: semiconductors,
thin metals, biomembranes, crystal oscillators, very long-term temperature
variations, the loudness of orchestral music, fluctuations in the
Earth's rotation speed, fluctuations in the intensity of cosmic rays,
heartbeats, postural control, magnetoencephalography and electroencephalography
in the brain, etc.\cite{Milotti2002,Bosnian2001}. 

There have been many discussions about the origin of pink noise\cite{Milotti2002,Bosnian2001,Hooge1994},
but there seems to be no clear conclusion. Many models have been proposed
that give rise to pink noise, but no universal mechanisms have been
discovered. 

Since pink noise is ubiquitous, the mechanism should be simple enough.
However, all the applications of the basic concepts and techniques
of the standard statistical mechanics seem to have encountered conflicts
and disputes. Then people have tended to consider more fundamental
concepts that can rewrite the theory of standard statistical mechanics. 

A typical mechanism for producing arbitrary low-frequency fluctuations
would be the wave beat, or amplitude modulation, of the primary high-frequency
fluctuations. This amplitude modulation would be successful for pink
noise if the frequencies were more concentrated in a small range.
Then the secondary beat wave can have lower frequencies. One of the
authors has already proposed this mechanism for the pink noise of
sounds and music\cite{Morikawa2021}. 

Furthermore, this concentration should be cooperative and systematic
to form the power-law PSD. We propose at least three types of cooperative
systems that can produce pink noise. They are a) synchronization (section
three), b) resonance (section four), and c) the infrared (IR) divergence
(section five). 

If the pink noise were an amplitude modulation, the demodulation mechanism
should also exist. This is because the entire modulated data has only
high-frequency information, while the data after demodulation can
explicitly show the low-frequency information, including the pink
noise. The demodulation mechanism can be intrinsic to the system or
it can be prepared in the measurement procedure. Many demodulation
mechanisms make the pink noise phenomena diverse: taking the square
of the original signal, rectification, thresholding, etc. For example,
when the electrical current or voltage exceeds the threshold in the
biological body, ignition occurs and produces spikes in the nerve
cells. Thus the possible pink noise in the electric current is transferred
to the nerve signal. 

We begin our discussion in section two, listing crucial clues to the
origin of pink noise, all of which point to the possibility that pink
noise is amplitude modulation. We then propose three mechanisms that
lead to the modulation. In section three, we discuss the most typical
mechanism synchronization. We show that a) exponential synchronization
yields a power index of $-1,$ and power-law synchronization yields
a power index slightly different from $-1$. In section four, b) resonance
also yields pink noise since the concentration of the excited eigenmodes
around the fiducial frequency is systematically approximated by the
exponential function in the relevant domain. In section five, c) infrared
divergence in the bremsstrahlung can give pink noise. In section six,
we discuss the robustness of pink noise and several demodulation mechanisms
that yield a variety of pink noise. In the final conclusion section
seven, we summarize our proposal and possible verifications based
on the points presented in section two. We also summarize our future
prospects of amplitude modulation on a variety of systems. 

\section{Some crucial clues for pink noise \label{sec:Several-crucial-hints}}

We will now list some crucial clues to the origin of the pink noise.
This process is quite important, because it can clarify which principles
of statistical mechanics are useful and which are not useful to describe
the pink noise. 
\begin{enumerate}
\item Wave \\
Systems that exhibit pink noise are often waves: sound waves, electric
current, air-fluid, liquid flow, etc. Waves can interfere with each
other. Thus the interference of waves can be a clue to get pink noise. 
\item Small system and seemingly long memory\\
It is bizarre that an ultra-low frequency signal can come from a very
small system. As an extreme example\cite{Liu2013}, the semiconductor
films of 2.5nm layers give observable pink noise. A small semiconductor
can have pink noise down to $10^{-7}\mathrm{Hz}$\cite{Dukelov1974},
and voltage fluctuations through a semiconductor show pink noise from
about $1\mathrm{Hz}$ to $10^{-6.3}\mathrm{Hz}$\cite{Caloyannides1974}.
These remarkable low frequencies sound almost impossible for ordinary
small systems. In this context, if the Wiener-Khinchin theorem $S(\omega)=\int_{-\infty}^{\infty}d\tau\int_{-\infty}^{\infty}dt\langle x(t)x(t-\tau)\rangle e^{-2\pi i\omega\tau}$
were correct, then the strong low-frequency signal in $S(\omega)$
of the pink noise would necessarily indicate the non-vanishing long-time
correlation $\langle x(t)x(t-\tau)\rangle$. Therefore, the Wiener-Khinchin
theorem may not hold for pink noise. 
\item Apparent no lower cutoff in the PSD\\
It is often discussed that the pink noise does not seem to have an
explicit lower cutoff in the PSD determined by any physics governing
the system. Therefore, the system exhibiting pink noise may not be
in a stationary state. Therefore, it may be useless to have discussions
based on the stationarity of the system. 
\item Independence from dissipation\\
It is remarkable that the pink noise appears even in the Hamiltonian
mean-field (HMF) model, which is a strictly conservative system \cite{Yamaguchi2018}
and has nothing to do with dissipation. Thus the usual fluctuation-dissipation
theorem of the type $\left\langle \delta x^{2}\right\rangle \propto RkT$
may not hold for the pink noise($R$ is the electric resistance and
$kT$ is the temperature). 
\item Square of the original signal\\
When deriving the pink noise, it is often the case that the original
time sequence is squared prior to the PSD analysis. For example, in
the case of music\cite{Voss1977}, the sound wave data should always
be squared for PSD; the authors claim that this squared data is the
loudness. Similarly, in the case of the HMF model\cite{Yamaguchi2018},
the authors always take a square of the original variables in order
to obtain the pink noise. In both cases, the original data before
taking the square does not show any pink noise. In the case of the
electric current, this procedure is not manifest, although the seminal
paper \cite{Johnson1925} emphasizes the square of voltage $V^{2}$
for PSD. 
\end{enumerate}
From the above five clues, we speculate that the beets of many synchronized
waves may be the origin of 1/f noise. A simple superposition of two
waves $\sin(\omega t+\lambda t)+\sin(\omega t-\lambda t)=2\cos(\lambda t)\sin(\omega t)$
with $\omega\gg\lambda>0$ has no low frequency component around $\lambda$
in the PSD. On the other hand, the square of the superposed wave above
has a low-frequency signal, \textit{i.e.,} the beats, around $2\lambda$
in its PSD. Incidentally, it is sometimes confusing that the wave
beat is ''audible'' although the PSD of the original superposition
of the two waves does not show the corresponding low-frequency signal. 

The above argument reminds us of a typical musical instrument, the
Theremin \cite{Glinsky2000}, which uses the wave beat. By mixing
the high-frequency signals of 1000kHz and 999.560kHz generated by
an electric circuit, the low-frequency signal of 440Hz can be extracted
as audible sound. The latter frequency can be varied slightly by the
player's hand, antenna distance, and capacitance, to produce the desired
frequency signal. Thus the amplitude modulation can produce arbitrary
low-frequency signals within a small-size system. The modulated signal
has no intrinsic memory and has nothing to do with dissipation. 

Another familiar device is the AM radio which clearly shows the wave
beat or amplitude modulation (AM). By using 526.5kHz to 1606.5kHz
radio waves, the low-frequency audible signal is extracted. In this
case, the rectification (demodulation) process is essential to obtain
audible low-frequency signals. This demodulation process is also essential
for the pink noise in our proposal. In later sections, we will see
a variety of pink noises in the many ways to demodulate. 

The above five points will also be an elementary verification of our
proposal. This will be discussed in later sections. 

There appear to be several causes of the wave beat that forms pink
noise, but the concentration of the wave frequencies is the essence
of low-frequency signals. We will now focus on such causes separately
in the following sections: a) cooperative waves, b) resonance, and
c) infrared divergence. 

\section{Beats from cooperative waves\label{sec:Beats-from-cooperative}}

In this section, we will analyze the cause of wave beat, especially
when the frequencies of the waves spontaneously approach with each
other. We consider cooperative systems that exhibit this behavior. 

\subsection{Exponential Approach\label{subsec:an-exponential-approach}}

The most typical type of synchronization would be the exponential
approach, such as in the case of the Kuramoto model \cite{Kuramoto1975},
$\omega=e^{-\lambda t}$ where $\omega$ is the frequency and $\lambda$
is the approach speed, and $t$ is the time. Then the frequency distribution
function $P(\omega)$ and the time distribution function $p(t)$ are
related to each other by $P(\omega)d\omega=p(t)dt$. If we assume
the stationarity of the fluctuation, we set $p(t)\equiv p=const$.
Then, 

\begin{equation}
P(\omega)=p(t)|d\omega/dt|^{-1}=p\lambda^{-1}\omega^{-1}\propto\omega^{-1}.\label{eq:1}
\end{equation}
\label{subsec:an-exponential-approach-1} It is interesting that the
exponential function gives the power index exactly $-1$. 

\begin{figure}

\includegraphics[width=8cm]{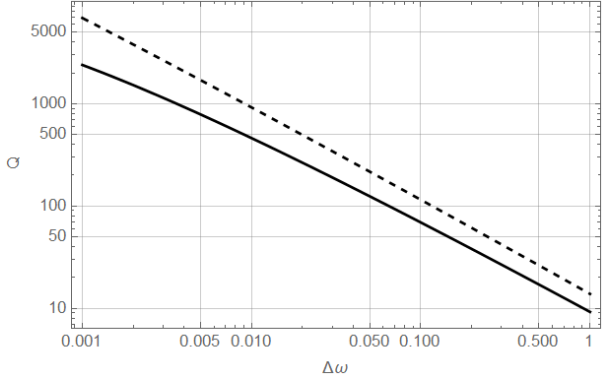}\caption{Examples of $Q(\Delta\omega)$ for the cases $p=1,\lambda=1,\omega_{2}=10^{5}$
and $\omega_{1}=10^{-4},10^{-6}$ (solid and dashed curves, respectively).
The detailed behavior of $Q(\Delta\omega)$ depends on the upper and
lower bounds of the integration. }
\label{fig1}
\end{figure}

The observed beat is the interference of the pair of frequency distributions
above, and the beat frequency $\Delta\omega$ has its probability
distribution function $Q(\Delta\omega)$ as 

\begin{equation}
\begin{aligned}Q(\Delta\omega) & =\int_{\omega_{1}}^{\omega_{2}}d\omega P(\omega+\Delta\omega)P(\omega)\\
 & =\frac{p^{2}}{\lambda^{2}\Delta\omega}\ln\left[\frac{\omega_{2}\left(\omega_{1}+\Delta\omega\right)}{\omega_{1}\left(\omega_{2}+\Delta\omega\right)}\right]\\
\\
\end{aligned}
\label{eq:2}
\end{equation}
which again is proportional to $\left(\Delta\omega\right)^{-1}$with
small modification factor of $\ln[...\Delta\omega]$. The detail of
the full form $Q(\Delta\omega)$ depends on the boundaries of the
integration domain $\omega_{1}<\Delta\omega<\omega_{2}$. Typical
examples are shown in Fig.\ref{fig1}.

\begin{figure}

\includegraphics[width=8cm]{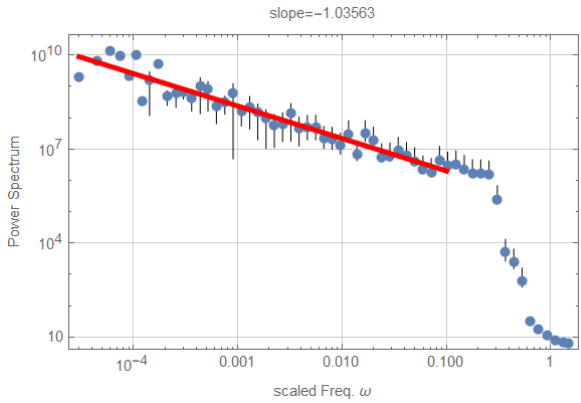}\caption{The PSD of $\phi^{2}$ is shown with $\omega=10$, $c=0.2$, and $r$
is a random field in the range {[}0,30{]}. 1000 sine waves are superimposed
according to Eq.\ref{eq:3} The power index can change up to about
$0.1$ for each run. This PDF shows the pink noise of index -1 for
four decades. }
\label{fig2}

\end{figure}

The pink noise is robust, and the frequency distribution is directly
reflected in the PDF of the waves at those frequencies, 

\begin{equation}
\phi\left(t\right)=\sum_{i}\sin\left(2\pi\omega(1+ce^{-r_{i}})t\right),\label{eq:3}
\end{equation}
where $\omega$ is a fiducial frequency, $c$ is a mixing constant,
$r_{i}$ the Poisson random variable in some range for each sinusoid
and $i$ runs from $1$ to some upper limit. This is demonstrated
in Fig. \ref{fig2} where the PSD of $\phi^{2}$ is shown. 

\begin{figure}
\includegraphics[width=8cm]{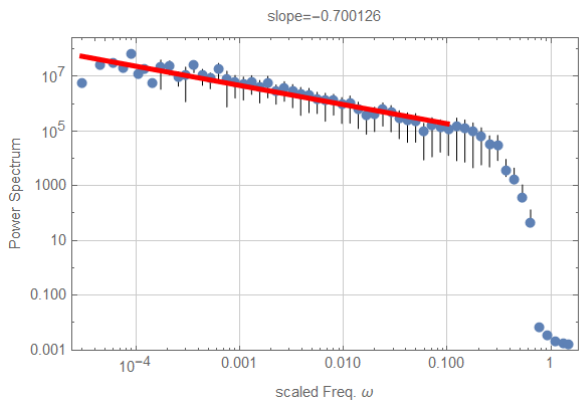}\caption{Same as Fig. \ref{fig2}, but the sine waves are superimposed with
random phase $\theta_{i}$ for each:$\sin\left(2\pi\omega(1+ce^{-r_{i}})t+\theta_{i}\right)$.
The power index drops a bit to $-0.7$, but this PDF shows the robustness
of the pink noise from the wave beat. }
\label{fig3}

\end{figure}

\textsubscript{}The pink noise is robust, and the randomization of
each phase of the sin-wave does not change the PDF except that the
power index is slightly reduced, as shown in Fig.\ref{fig3}. 

\begin{figure}
\includegraphics[width=8cm]{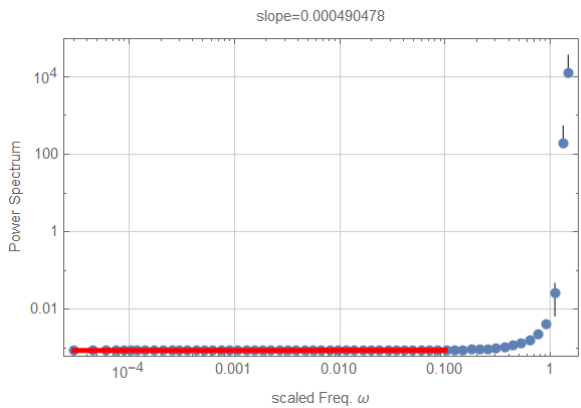}\caption{Same as Fig.\ref{fig2}, but this is PDF for the original signal $\phi$.
Pink noise never appears in this case, indicating that the noise arises
from the wave beat. }

\label{fig4}
\end{figure}

It is essential that the square of the signal $\phi^{2}$ does show
pink noise in PSD as in Fig. \ref{fig1} while the original signal
itself $\phi$ does not show any feature at low-frequency region as
shown in Fig.\ref{fig4}. This fact manifestly demonstrates the pink
noise comes from the wave beat. 

\subsection{Power Approach \label{subsec:power-approach}}

\begin{figure}

\includegraphics[width=8cm]{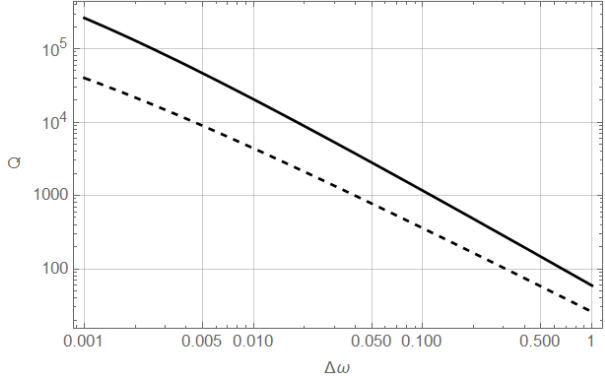}\caption{Examples of $Q(\Delta\omega)$ for the cases $p=1,\lambda=1,\omega_{2}=10^{5}$
and $\omega_{1}=10^{-4},c=1,\beta=1.2$and $1.33$ (solid and dashed
curves, respectively).}
\label{fig5}
\end{figure}

Another popular type of synchronization would be the power approach
$\omega=t^{-\alpha}$. Repeating the same calculations as above, we
obtain the frequency distribution function as

\begin{equation}
\ensuremath{P(\omega)=\underbrace{p(t)}_{p\text{ const. }}|d\omega/dt|^{-1}=c\omega^{-\beta}}
\end{equation}
where $c\equiv p\alpha^{-1},\beta\equiv\left(1+\frac{1}{\alpha}\right).$
The probability distribution function $Q(\Delta w)$ of the beat
frequency $Q(\Delta w)$ is given by 

\begin{equation}
\ensuremath{Q(\Delta\omega)=\int_{\omega_{1}}^{\omega_{2}}d\omega P(\omega+\Delta\omega)P(\omega)}
\end{equation}
Then,
\begin{align}
Q(\Delta\omega) & =\frac{1}{\Delta\omega\left(1-\beta\right)}\nonumber \\
 & \left[c^{2}\omega{}^{1-\beta}(\Delta\omega+\omega)^{1-\beta}{}_{2}F_{1}\left(1,2-2\beta;2-\beta;-\frac{\omega}{\Delta\omega}\right)\right]_{\omega_{1}}^{\omega_{2}}\\
 & \propto\Delta\omega{}^{-1-\left(2/\alpha\right)},\nonumber 
\end{align}
if we expand with respect to small $\omega_{1}$ and small $\Delta\omega.$
The exponent is less than $-1$ for $\alpha>0$, and greater than
$-1$ for $\alpha<0$ but the fiducial power is $-1$. Typical examples
are shown in Fig.\ref{fig5}. 

\begin{figure}
\includegraphics[width=8cm]{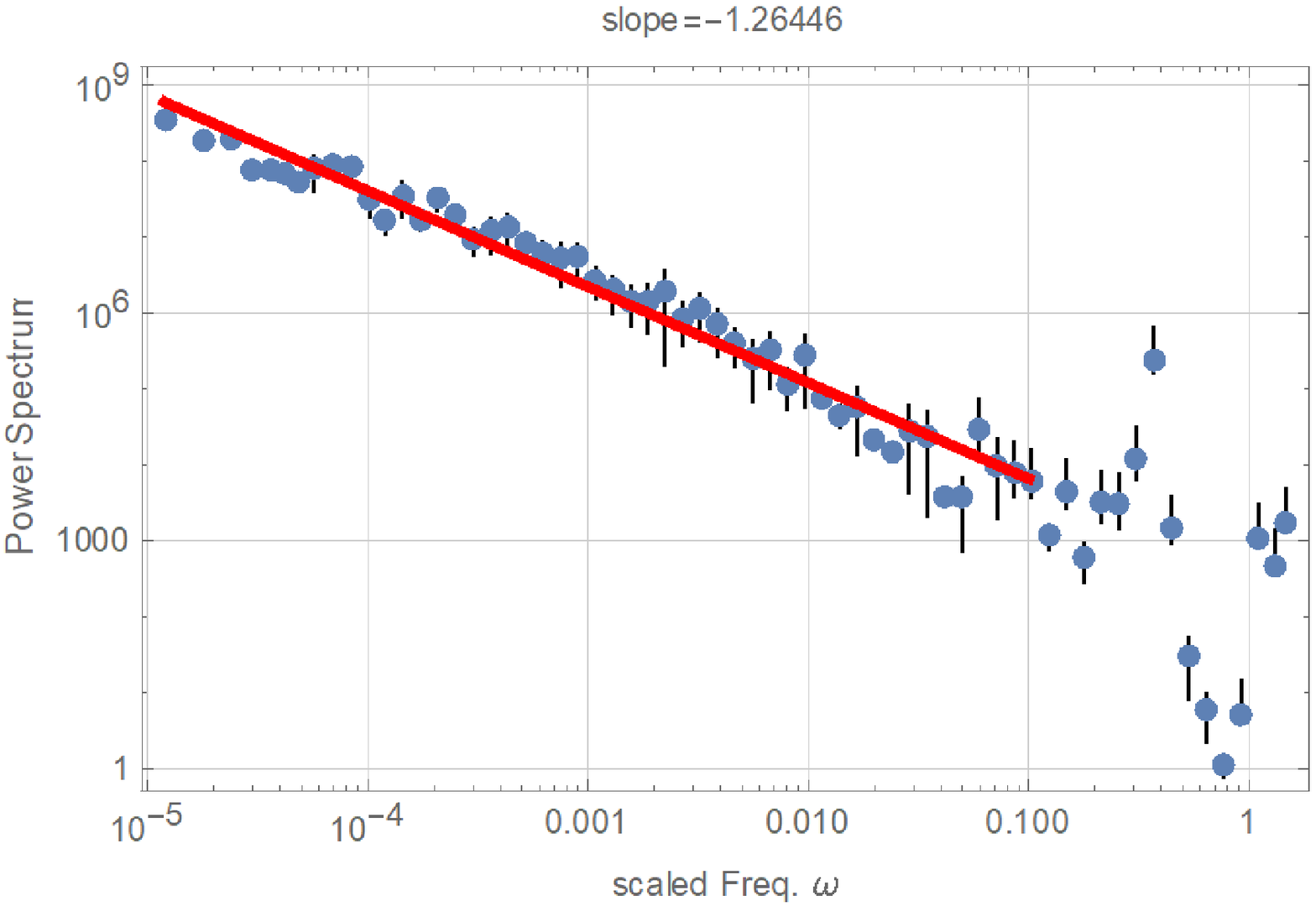}\caption{The PSD is shown for $\phi^{2}$ with $\alpha=3,\omega=440$, $c=0.3$,
and $r_{i}$ is a random field in the range {[}0,20{]}. 200 sine waves
are superimposed according to Eq.\ref{eq:8} This PDF shows the pink
noise of index -1.3 for four decades. }
\label{fig6}

\end{figure}

\begin{figure}

\includegraphics[width=8cm]{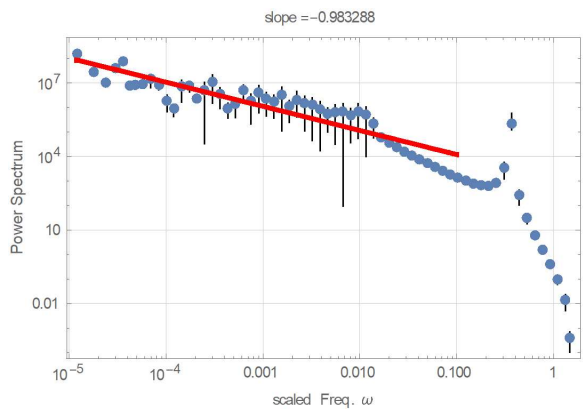}\caption{The PSD is shown for $\phi^{2}$ with $\alpha=-3,\omega=440$, $c=0.01$,
and $r_{i}$ is a random field in the range {[}0,1{]}. 200 sine waves
are superimposed according to Eq.\ref{eq:8} This PDF shows the pink
noise of index -1 for three decades. }
\label{fig7}

\end{figure}

A typical wave signal can be constructed as before,

\begin{equation}
\phi=\sum_{i}\sin\left(2\pi\omega(1+cr_{i}^{-\alpha})t\right),\label{eq:8}
\end{equation}
and PSD for $\phi^{2}$ are shown in Fig.\ref{fig6} for $\alpha=3$,
and in Fig. \ref{fig7} for $\alpha=-3.$

Although the above demonstrations are typical simple models of the
cooperative waves, the frequencies are fixed. However, it is also
possible to consider dynamical cooperative systems with time-dependent
frequencies, and they often show pink noise; macroscopic coupled spin
models \cite{Nakamichi2012} and the Hamiltonian mean field model
\cite{Yamaguchi2018}. Since the discussion of these is beyond the
scope of this paper, we will cover them in a separate paper soon. 

\section{Beats from Resonance\label{sec:Beats-from-Resonance}}

We now consider the resonance, which produces the spontaneous concentration
of frequencies and the wave beats. When the system with the intrinsic
eigenfrequency $\Omega$ is stimulated (repeatedly), it emits the
wave mode of the frequency $\Omega$ as well as those close to $\Omega$.
Resonance thus ensures the concentration of frequencies in a small
range. Since these frequencies are close to each other, the waves
of these frequencies beat and produce a signal in low-frequency regions. 

Suppose a typical case of the resonance characterized by the resonance
curve, the Cauchy distribution
\begin{equation}
R[\omega]=\frac{1}{\left(\frac{\kappa}{2}\right)^{2}+\left(\omega-\Omega\right){}^{2}},
\end{equation}
where $\Omega$ is the resonance frequency and $\kappa$ characterizes
the sharpness of the resonance. We will interpret that this function
$R[\omega]$ as proportional to the number of $\omega$-modes in the
resonator. Then the frequency distribution function $P(\omega)$ is
given by the inverse function of $R[\omega]$, as 
\begin{equation}
\omega=R^{-1}[t]=\frac{\sqrt{-t\left(\kappa^{2}t-4\right)}}{2t}+\Omega,\label{eq:10}
\end{equation}
where we have chosen the upper half of the inverse of $R[\omega]$,
since the lower half is symmetric to the upper half. 

\begin{figure}

\includegraphics[width=8cm]{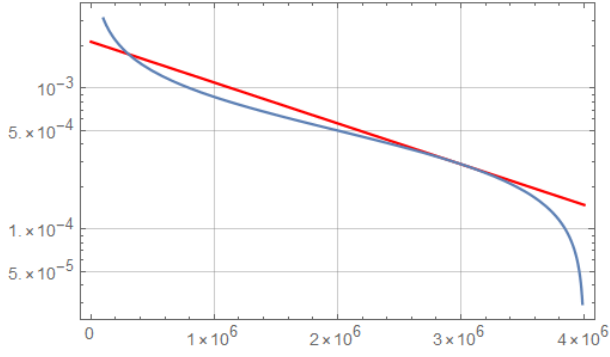}\caption{Demonstration of Eq.\ref{eq:10} in the log-linear graph. The function
$\omega(t)$ can be approximated by the exponential function (red
straight line) with the same inclination at the inflection point of
$\omega(t)$, especially in the large-t range that is relevant for
the low-frequency beats. }
\label{fig8}
\end{figure}

It is possible to make a naive approximation of Eq.\ref{eq:10} by
the exponential function $\omega=Ae^{-Bt}$, where the constants $A,B$
are determined at the inflection point of Eq.\ref{eq:10}, as shown
in Fig.\ref{fig8}. We already know that this exponential function
gives the exact pink noise of slope $-1$ in PSD. 

\begin{figure}

\includegraphics[width=8cm]{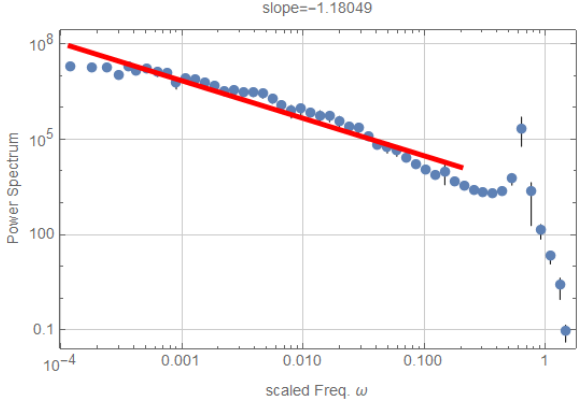}\caption{PSD of the time sequence $\phi(t)^{2}$ generated by Eq.(\ref{eq:12})
with $\kappa=0.1,\Omega=10,$ and the domain of the random field $r_{i}$
is $[0,10]$. We have superimposed $100$ sine waves, and this PSD
shows the approximate power law of index $-1.2$.}

\label{fig9}
\end{figure}

This is demonstrated in Fig.\ref{fig9}, where the PSD is plotted
for the square $\phi(t)^{2}$of the time sequence $\phi(t)$ generated
by 

\begin{equation}
\phi(t)=\sum_{i}\sin\left(2\pi R^{-1}\left(r_{i}\right)t\right).\label{eq:12}
\end{equation}

However, the system analysis is not easy. Using the relation $P(\omega)d\omega=p(t)dt$
with $p(t)\equiv p=const$, we obtain the frequency distribution function
$P(\omega)$ as 
\begin{equation}
\ensuremath{P(\omega)=p|d\omega/dt|^{-1}=\frac{32p(\omega-\Omega)}{\left(\kappa^{2}+4(\omega-\Omega)^{2}\right)^{2}}},
\end{equation}
which cannot be reduced to a single power form if $\kappa$ is finite. 

Further complications arise from the actual resonant system, which
has complicated overtones and multiple eigenfrequencies that systematically
contribute to the pink noise. A fully systematic derivation of pink
noise for each concrete resonant system requires further investigation.
Since this is beyond the scope of this paper, we do not discuss it
further here, but it will be analyzed in a separate paper soon. 

\section{Beats from IR divergence}

We now consider the third cause of the spontaneous concentration of
frequencies from the infrared divergence. This class of systems exhibiting
pink noise is quite diverse, but can be reduced to the system composed
of electrons and photons described by electrodynamics. 

In this context, a quantum origin of pink noise was once proposed
by using quantum interference\cite{Handel1975,Handel1980}. It claims
that the scattered electron state, after emission of a photon of frequency
$\omega$, and the unscattered electron state interfere with each
other to produce a beat of frequency $\omega$. However, this theory
has been criticized \cite{Kiss1986,Nieuwenhuizen1987}, mainly because
quantum interference does not really occur; the scattered and non-scattered
states are orthogonal to each other and have no chance to interfere.
Even the introduction of the coherent state basis does not work. Incidentally,
some other criticisms are not valid. 

The essence of the pink noise is not the quantum interference, but
the back-reaction of the emission of massless particles to the classical
current and the classical wave beat interference. In this paper, we
focus on such a classical description of electromagnetism. 

In the semiconductor, the electric current can be classical beyond
the scale of the free streaming length, about $10nm$, which is several
tends of the lattice size. When the system size is about $1mm$, there
are $10^{10}$ classical current elements. When the electron meets
any impurity, it changes its momentum from $p^{i}$ to $p^{f}$, emitting
the photon of momentum $p^{i}-p^{f}.$ Starting with the classical
current, 

\begin{equation}
j_{\mu}(x)==-\frac{ie}{(2\pi)^{4}}\int d^{4}ke^{-ik\cdot x}\left(\frac{p_{\mu}^{i}}{p^{i}\cdot k}-\frac{p_{\mu}^{f}}{p^{f}\cdot k}\right),
\end{equation}
the number of emitted photon is given by 

\begin{equation}
dN=e^{2}\left|\frac{\varepsilon\cdot p^{i}}{k\cdot p^{i}}-\frac{\varepsilon\cdot p^{f}}{k\cdot p^{f}}\right|^{2}\frac{d^{3}k}{2(2\pi)^{3}k^{0}},
\end{equation}
which is IR divergent \cite{Itzykson1980}, and $\varepsilon$ is
the polarization vector. 

We assume that the average classical electric current has a fiducial
frequency $\Omega$, which is determined by the applied voltage and
the conductivity before the interaction with the impurities. Each
scattering with the impurities emits light of energy $\omega$ and
exerts a back-reaction on the current of amount energy-shift $\omega$
with the probability $1/\omega$ (bremsstrahlung). Then the original
current cascades into the superposition of an enormous number of local
currents with frequencies $\Omega-\omega_{i},i=1,2,...N$. Then each
pair of these currents makes beats with all the possible differences
$\left(\Omega-\omega_{i}\right)-\left(\Omega-\omega_{j}\right)=\omega_{j}-\omega_{i},1\leqq i,j\leqq N.$
This process is the same as the previous case of the exponential approach
\ref{subsec:an-exponential-approach}, and many classical currents
with slightly different frequencies interfere to give the wave beat
as in Eq.(\ref{eq:2}) and thus the pink noise as in Fig.\ref{fig2}. 

In any case, quantum interference is not the essence of the origin
of pink noise, but the classical synchronized waves are crucial. In
this context, the coherent dressed state formalism for QED was developed
to cancel the infrared divergence associated with the massless photon.
This theory is well summarized in\cite{Hirai2019,Hirai2021}, although
most authors assume (semi-)classical background currents ab initio,
and the classical degrees of freedom are not correctly derived. 

The derivation of the classical degrees of freedom in QED is possible
in the closed time-path formalism of the effective action associated
with an unstable state. The IR divergence of the theory requires the
separation of the classical statistical kernel from the complex effective
action. Then the Langevin equation with classical noise is derived
from the effective action and can describe the classical evolution
of currents \cite{Morikawa2022}. 

This formalism requires a more systematic discussion than we can
give here. However, we will report this theory, including the classical-quantum
interference, in a separate paper. 

\section{Discussions}

So far, we have proposed three kinds of origin of the synchronizing
waves, which gives systematic beats and produce pink noise. Since
the pink noise is generated by the wave beat or the amplitude modulation,
any demodulation process is required for observation. This demodulation
process may be a)intrinsic mechanisms associated with the system or
b) operational processes associated with the data reduction for PSD.
In either case, the demodulation process provides robustness and a
variety of pink noise. This section is devoted to showing some examples
of such robustness and variety. 
\begin{enumerate}
\item \emph{fiducial}: The fiducial signal is the one discussed in \ref{subsec:an-exponential-approach},
with the same parameters of Fig.\ref{fig2}: $\omega=10$, $c=0.2$,
and $r_{i}$ is a random field in the range {[}0,30{]}. There, $10^{3}$
sinusoids are superimposed according to Eq.\ref{eq:8}. The squared
signal $\phi^{2}$shows a clear pink noise of slope $-1.0$ as in
Fig.\ref{fig2}. 
\item the threshold for $\phi^{2}$: We set the new data zero for the $\phi^{2}$
data that is \emph{smaller than the mean} and leave the other data
\emph{as they are}. The PSD shows pink noise with slope -1.0, almost
no change from the fiducial case. This case may apply to the nerve
system, where only a voltage greater than some threshold can produce
a spike signal. 
\item on-off threshold for $\phi^{2}$: We set the new data zero for the
$\phi^{2}$ data that is \emph{less than the mean} and set the other
data \emph{to 1}. The PSD shows pink noise with a slope of -0.94. 
\item on-off inverse threshold for $\phi^{2}$: This is \emph{the opposite
of case 3}. We set the value 1 for the $\phi^{2}$ data that is smaller
than the mean and set the other $\phi^{2}$data to 0. The PSD shows
pink noise with a slope of -0.94, exactly the same as in case 3. 
\item threshold for original\emph{ data $\phi$:} we set the new data zero
for the $\phi$ data that is smaller than the mean and set the other
data as is. The PSD shows pink noise with a slope of -0.98. 
\item rectification of the original data $\phi$: We set the new data\emph{
to zero for the $\phi$ data that is negative} and leave the other
data as is. The PSD shows pink noise with a slope of -1.2. This may
apply to some electric circuits contaning transistors, diodes, and
vacuum tubes. 
\item sequence of locally averaged $\phi^{2}$: We divide the entire time
sequence of $\phi$ into\emph{ $10^{3}$ segments} and apply \emph{a
quadratic average in each} segment. The PSD shows pink noise with
a slope of -1.1. This is the data treatment in the original experiment
\cite{Johnson1925}. 
\item sequence of locally averaged $\phi$: S\emph{ame as case 7, but we
apply a simple} average in each segment. The PSD shows NO pink noise
at all, and the power is positive +0.8. 
\item coarse time resolution for $\phi^{2}$: We \emph{reduce the number
of sample points} to half of the original. The PSD shows an almost
pink noise with a slope of -1.1. 
\item fewer superimposed waves: We \emph{reduce the number of superimposed
waves} from the fiducial $10^{3}$ to $10$. The PSD shows NO pink
noise. 
\item more superimposed waves: We \emph{increase the number of superimposed
waves} from the fiducial $10^{3}$ to $10^{4}$. The PSD shows pink
noise with a power of -0.94. 
\item longer time sequence: We extend the \emph{time sequence from the fiducial
$10^{4}$ to }$10^{5}$. The PSD shows pink noise with a slope of
-1.0; the same as before, but with a power law extended by a decade. 
\item multiple fiducial frequencies; We changed the fiducial frequency from
the\emph{ original single to 5, randomly selected from 0 to 20}. The
PSD shows pink noise with a slope of -1.5. 
\end{enumerate}
As examined above, there are multiple demodulation processes. They
are classified as a) system-intrinsic and b) operational in the data
reduction, although the classification is not exclusive. Examples
of a) are thresholding and rectification: cases 3,4,5,6. Examples
of b) are data squaring: cases 1,2,7.

Cases 9,11,12,13 show some robustness of pink noise. 

\section{Conclusions and prospects}

We have discussed the origin of pink noise from the beat of cooperative
waves. We have examined three possible causes for this cooperative
effect: synchronization, resonance, and IR divergence. There may be
more mechanisms. We point out the verifiability/falsifiability of
our model based on the five crucial observations for the pink noise
in section \ref{sec:Several-crucial-hints}.
\begin{enumerate}
\item Wave \\
The wave is essential for producing beat and amplitude modulation.
The wave may be hidden inside the system, and the data may be obtained
after it passes through the threshold. If we cannot find a coherent
wave in the system, our model cannot be applied. 
\item Small system and apparent long memory\\
Although the amplitude-modulated fluctuation, the primary fluctuation,
may accept the Wiener-Khinchin theorem, the demodulated fluctuation,
the secondary fluctuation, does not accept the theorem because the
secondary fluctuation does not appear in the PSD before any demodulation
process. If we find the successful Wiener-Khinchin theorem for pink
noise, our model cannot be applied. 
\item Apparent no lower cutoff in the PSD\\
The beat of the cooperative wave or the amplitude modulation can yield
an infinitely low-frequency signal from inside a finite system within
the observational constraints. Therefore, if an intrinsic lower-cutoff
frequency is found in the pink noise, our model cannot be applied.
\item Independence from dissipation\\
The beat of the cooperative wave or the amplitude modulation is a
secondary fluctuation caused by wave synthesis. Therefore, the dissipation
may destroy the pink noise because it may cancel the fragil wave beats. 
\item Square of the original signal (necessity of the demodulation process)\\
The amplitude modulation needs some demodulation process for observation.
The primary fluctuations before the demodulation do not appear in
the PSD. Our model for pink noise predicts the demodulation process
as either a)intrinsic to the system or b)operational in the data reduction.
If the demodulation is found in the system of pink noise, and the
pink noise disappears when the demodulation process is removed, then
our model is strongly favored. 
\end{enumerate}
Although we have proposed a basic model of pink noise, we still have
many problems with elaborating the present formalism. Some of them
have already been described in appropriate places with the keyword
'separate paper'. They are dynamical cooperative systems, actual resonant
systems, and systems with IR divergence. Among them, we summarize
the possibly resonant systems in Table \ref{tab:table1}. 

\begin{table*}
\caption{\label{tab:table1}List of systems that show pink noise possibly due
to the resonance effect, as we have discussed in \ref{sec:Beats-from-Resonance}.
This Table is preliminary, and the final analysis will be reported
in our papers soon. }
\begin{tabular}{lllll}
 & system & resonant mode & demodulation & description\tabularnewline
\hline 
1 & earthquakes & earth free oscillation\cite{Nishida2013} & fault rupture & 
\begin{minipage}[t]{0.3\textwidth}%
USGS World 30-year data shallower than 20km, magnitude 4-5 show pink
noise of slope -1.2 below 2.5 months.%
\end{minipage}
\tabularnewline
2 & icequakes & iceberg eigenfrequency & ice fault rupture & %
\begin{minipage}[t]{0.3\textwidth}%
NOAA Icequakes (Bloop) deep-sea sound\cite{NOAA2012} show pink noise
with slope -0.8. %
\end{minipage}\tabularnewline
3 & solar flare & five minute oscillation\cite{Thompson2018} & magnetic reconnection & %
\begin{minipage}[t]{0.3\textwidth}%
HESSI \cite{HESSI2022} solar flare luminosity curve for 16 years
shows pink noise with slope -0.9%
\end{minipage}\tabularnewline
4 & sunspots & same as above or macro spin model\cite{Nakamichi2012} & (intrinsic) & %
\begin{minipage}[t]{0.3\textwidth}%
Sunspot number time sequence from the year 1820 to 2010 shows pink
noise of slope -1.1 over the entire period\cite{Nakamichi2012}. %
\end{minipage}\tabularnewline
5 & $NO_{3}^{-}$ & same as above  & (intrinsic) & %
\begin{minipage}[t]{0.3\textwidth}%
$NO_{3}^{-}$ - Concentration during the years 1610\textendash 1904
in the DF01 antactica ice core\cite{Motizuki2022} shows pink noise
with slope -1.1.%
\end{minipage}\tabularnewline
6 & variable stars & same as above & (intrinsic) & %
\begin{minipage}[t]{0.3\textwidth}%
Some of the variable stars show pink noise. The light courve of Mira
(Red giant) for about six years \cite{AAVSO2022}shows pink noise
with slope -1.2. %
\end{minipage}\tabularnewline
7 & Suikinkutsu & 2-meter pottery cavity underground & data squared & %
\begin{minipage}[t]{0.3\textwidth}%
The water harp cave at HosenIn Kyoto shows pink noise of slope -0.8(left)
and -0.6(right) for about four decades.%
\end{minipage}\tabularnewline
8 & Big gong & eigenfrequency & data squared & %
\begin{minipage}[t]{0.3\textwidth}%
The big gong in Kyoto shows pink noise with slope -1.6. The small
gong shows NO pink noise. %
\end{minipage}\tabularnewline
\end{tabular}

\end{table*}

The list in Table \ref{tab:table1} is tentative and incomplete. It
will be completed in our future publications, including the verification
of our simple pink noise model. 

acknowledgments
We would like to acknowledge many valuable discussions with the members
of the Lunch-Time Remote Discussion Meeting, with the members of the
Department of Physics Ochanomizu University, with Manaya Matsui and
Izumi Uesaka at Kyoto-Sangyo University.

\end{document}